\documentclass[review]{elsarticle}
\usepackage{hyperref}
\hypersetup{colorlinks = true, allcolors = blue}
\usepackage[nameinlink,noabbrev]{cleveref}
\usepackage[margin=1in]{geometry}
\usepackage[font=footnotesize,labelfont=bf]{caption}
\usepackage{hhline}
\usepackage{verbatim}
\usepackage{xcolor}
\usepackage{multirow}

\biboptions{authoryear}

\begin{document}
\sloppy
\begin{frontmatter}

\title{\textbf{Requisites on viscoelasticity for exceptional points in passive elastodynamic metamaterials}}

\author[1]{Abhishek Gupta}
\author[1,2]{Ramathasan Thevamaran\corref{mycorrespondingauthor}}

\cortext[mycorrespondingauthor]{Corresponding author}
\ead{thevamaran@wisc.edu}

\address[1]{Department of Mechanical Engineering, University of Wisconsin-Madison, Madison, WI, 53706, USA}
\address[2]{Department of Engineering Physics, University of Wisconsin-Madison, Madison, WI, 53706, USA}

\begin{abstract}
The recent progress of non-Hermitian physics and the notion of exceptional point (EP) degeneracies in elastodynamics has led to the development of novel metamaterials for the control of elastic wave propagation, hypersensitive sensors, and actuators. The emergence of EPs in a Parity-Time symmetric system relies on judiciously engineered balanced gain and loss mechanisms. Creating gain requires complex circuits and amplification mechanisms, making engineering applications challenging. Here, we report strategies to achieve EPs in passive non-Hermitian elastodynamic systems with differential loss derived from viscoelastic materials. We compare different viscoelastic material models and show that the EP emerges only when the frequency-dependent loss-tangent of the viscoelastic material remains nearly constant in the frequency range of operation. Such type of loss tangent occurs in materials that undergo stress-relaxation over a broad spectrum of relaxation times, for example, materials that follow the Kelvin-Voigt fractional derivative (KVFD) model. Using dynamic mechanical analysis, we show that a few common viscoelastic elastomers such as Polydimethylsiloxane (PDMS) and polyurethane rubber follow the KVFD behavior such that the loss tangent becomes almost constant after a particular frequency. The material models we present and the demonstration of the potential of a widely available material system in creating EPs pave the way for developing non-Hermitian metamaterials with hypersensitivity to perturbations or enhanced emissivity.

\end{abstract}


\end{frontmatter}


\section{Introduction}

Using non-Hermitian singularities such as the exceptional points (EPs) to control wave dynamics is emerging as a new paradigm to develop novel elastodynamic \citep{dominguez2020environmentally,gupta2022reconfigurable,fang2021universal,lustig2019anomalous,rosa2021exceptional,hou2018tunable,shmuel2020linking,li2022experimental,hou2018p,wu2019asymmetric,fang2022emergence} and acoustic \citep{fleury2015invisible,thevamaran2019asymmetric,ding2018experimental,achilleos2017non,merkel2018unidirectional,zhu2014p,shi2016accessing} metamaterials with unusual functionalities like unidirectional invisibility \citep{zhu2014p,fleury2015invisible}, asymmetric mode switching \citep{doppler2016dynamically,elbaz2022encircling}, frequency-pure asymmetric transmission \citep{thevamaran2019asymmetric}, enhanced sensitivity \citep{kononchuk2022exceptional, rosa2021exceptional}, and enhanced emission \citep{gupta2022reconfigurable}. Unlike the geometric symmetries---e.g., rotational, translational, and periodicity---present in phononic crystals and acoustic metamaterials \citep{liu2000locally,boechler2011bifurcation,hussein2014dynamics,fang2006ultrasonic,matlack2016composite}, non-Hermitian systems utilize parity and time reversal symmetries ($PT$-symmetry) \citep{bender1998real} induced by either spatially distributed gain and loss mechanisms \citep{fang2021universal,rosa2021exceptional,fang2022emergence} or by energy partitioning between different wave propagation modes in purely elastic systems \citep{lustig2019anomalous,elbaz2022encircling}. 

While a plethora of research has studied EPs and their applications in PT-symmetric systems with balanced gain and loss, entirely passive (with no gain) systems with differential loss have also been shown to exhibit exceptional points \citep{shen2018synthetic,dominguez2020environmentally,thevamaran2019asymmetric,gupta2022reconfigurable,ferrier2022unveiling}. Loss or dissipation is often found as an intrinsic property of materials, whereas gain is induced often by pumping energy into the system from external sources via piezoelectric \citep{braghini2021non,wu2019asymmetric}, electroacoustic \citep{fleury2015invisible,shi2016accessing}, piezoacoustic \citep{hutson1961ultrasonic,gokhale2014phonon}, non-Foster circuits \citep{schindler2012symmetric}, and electromagnets \citep{bender2013observation}. Creating gain needs sophisticated, active, positive feedback control circuits, making systems energy expensive, bulkier, and difficult to control each active elements \citep{tsoy2017coupled,wu2019asymmetric,schindler2012symmetric,bender2013observation}. Realizing EPs in passive systems can make implementation more straightforward and integration into devices and structures seamless for engineering applications.

\begin{figure}[t]
	\centering
	\includegraphics[width=1\textwidth]{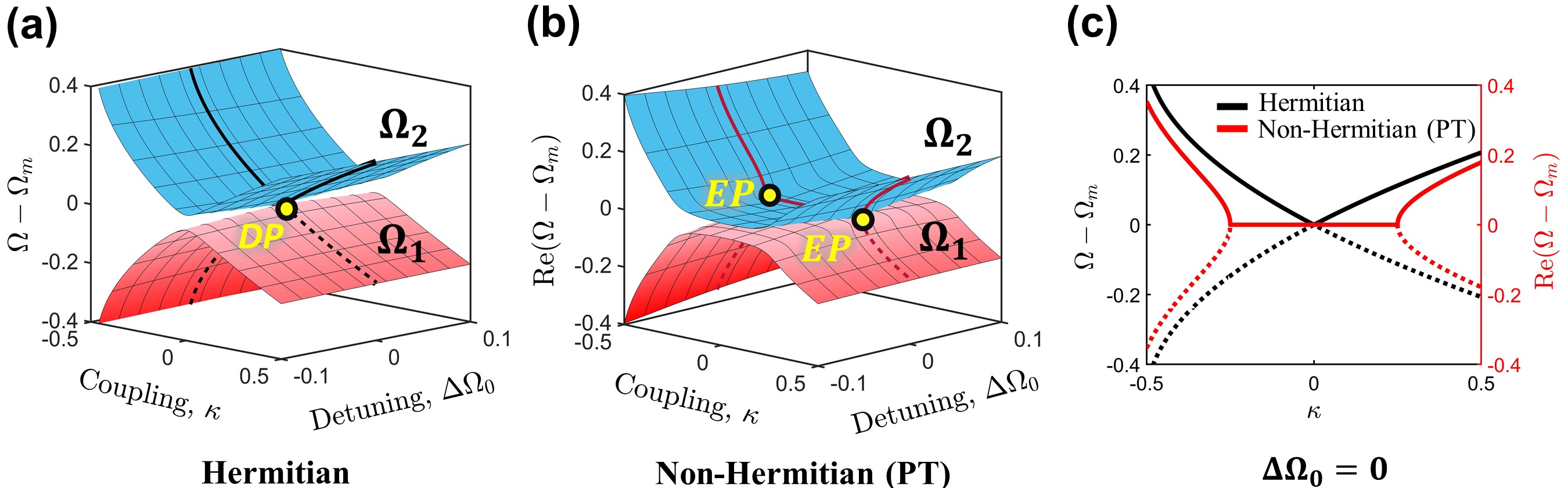}	\caption{(a) Linear bifurcation of eigenvalues $\Omega_1$ and $\Omega_2$ at a diabolical point in a coupled resonators system described by a $2\times2$ Hermitian Hamiltonian. Where, $\kappa$ is the coupling between the resonators, $\Delta\Omega_0$ is the frequency detuning between the resonators, and $\Omega_m$ is the mean of natural frequencies $\left(\Omega_m = {{\Omega_1+\Omega_2}\over 2}  \right)$. (b)  Square root bifurcation of eigenvalues at an exceptional point in a non-Hermitian coupled oscillators system with PT-symmetry. (c) Comparison of bifurcation diagrams between Hermitian and Non-Hermitian (PT) for $\Delta \Omega_0=0$. Introducing gain and loss in Hermitian system results in splitting of one DP into two EPs.}
	\label{her1}
\end{figure}

EPs are branch point singularities in the parameter space of a physical system where eigenvalues and corresponding eigenvectors coalesce and become degenerate \citep{el2018non,bender2019pt}. In contrast to degeneracy points in Hermitian Hamiltonians (diabolic points), where the bifurcation is linear, EPs exhibit higher-order bifurcations \citep{chen2017exceptional,hodaei2017enhanced} (\Cref{her1}). Any perturbation in the vicinity of an exceptional point results in a bifurcation of degenerate eigenvalues in the orthogonal direction (orthogonal bifurcation), making EPs very sensitive to external interference. The sensitivity of EPs has been exploited in developing hypersensitive gyroscopes \citep{wiersig2014enhancing,ren2017ultrasensitive,de2019high}, nanoparticle sensors \citep{chen2017exceptional,hodaei2017enhanced,rosa2021exceptional}, and accelerometers \citep{kononchuk2022exceptional,kononchuk2020orientation}. However, achieving a similar level of sensitivity in passive non-Hermitian systems is challenging regardless of their other utility such as in enhancing emissivity at the proximity to the EPs \citep{gupta2022reconfigurable}. Unlike sharp orthogonal EP phase transition in PT-symmetric systems, bifurcations in passive non-Hermitian systems exhibit an approximate transition with an avoided crossing of eigenmodes \citep{dominguez2020environmentally,bender2013twofold,gu2021controlling}. 

In this article, we report material design considerations for realizing EPs with sharp bifurcation in a passive non-Hermitian metamaterial in elastodynamic framework. Such a metamaterial can be realized by constituent elements containing a coupled mechanical oscillators (dimer) having differential loss. Incorporating a resonant element with a viscoelastic material for the dissipative component while the other resonator being an elastic conservative element will allow introducing non-Hermiticity to the metamaterial. While numerous plastics, rubbers, and foams exhibit viscoelasticity that can be described by different viscoelastic models, not every viscoelastic material will lead to a non-Hermitian metamaterial with a clear EP. The inherent-coupling between the storage and loss moduli (i.e. the real and imaginary parts of the complex dynamic modulus) by the Kramers-Kronig relation \citep{lakes2017viscoelastic,findley2013creep} makes the experimental realization a significant challenge and require a judicial design \citep{gupta2022reconfigurable}. We investigate different viscoelastic material models in comparison to a coupled oscillator model consisting of differential frictional damping and formulate requisites on frequency-dependent loss tangent to achieve a sharp EP bifurcation that is critical for exploiting its hypersensitivity to system parameters. Our findings will guide the experiments \citep{gupta2022reconfigurable} on achieving sharp EPs in systems containing viscoelastic materials to create non-Hermiticity. Because the governing non-Hermitian Hamiltonian of our system is mathematically similar to all other bimodal physical systems, our results can be applied broadly to other fields of physics as well.

This article has the following structure: We first discuss general rate independent frictional (hysteretic) damping and compare the occurrence of EPs in a coupled oscillators dimer for PT-symmetric case with balanced gain and loss and passive non-Hermitian case with differential loss. In the following sections, we revisit the basics of viscoelastic materials to test the feasibility of using them in creating non-Hermiticity. We individually analyze three different viscoelastic material models to identify the best suited viscoelastic material to create a passive elastodynamic metamaterial that support the formation of an EP. Finally, we report experimentally measured dynamic properties of viscoelastic elastomers---polydimethylsiloxane (PDMS), polyurethane rubber, and natural rubber and demonstrate a concept experimental design to achieve EPs. 

\begin{figure}[t]
	\centering
	\includegraphics[width=\textwidth]{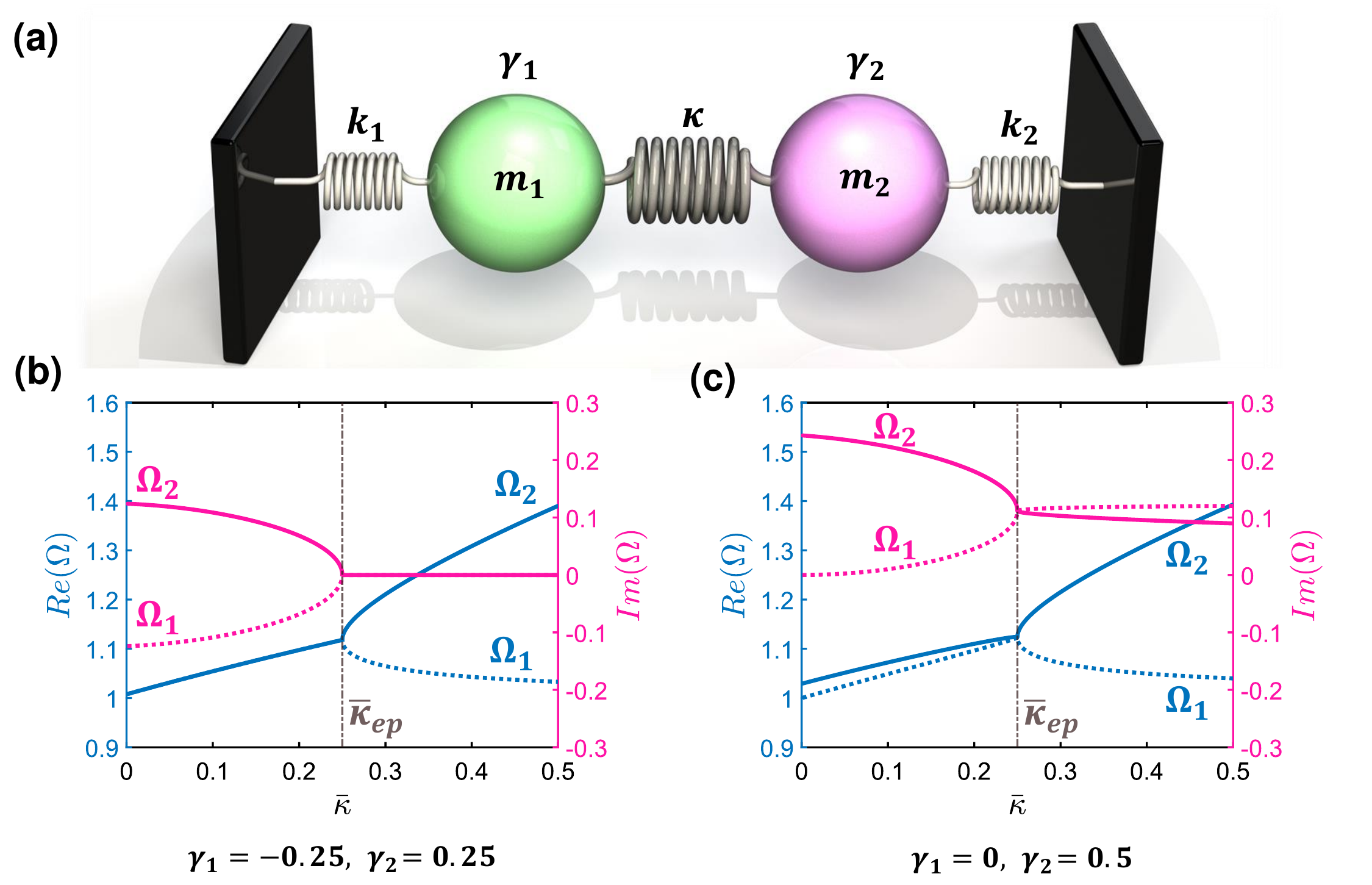}
	\caption{(a) A coupled oscillator dimer with differential damping, the green oscillator is less damped than the purple oscillator. (b) Real and imaginary parts of natural frequencies for balanced gain and loss (PT symmetric). (c) Real and imaginary parts of natural frequencies for noLoss-Loss differential damping case (passive non-Hermitian).}
	\label{her2}
\end{figure}

\section{Coupled oscillators with differential frictional damping}

Discrete coupled oscillators exhibit two individual vibration modes, one in-phase and one out-of-phase. For Hermitian systems, the resonant frequencies of these modes are real and distinct. However, these modes begin interacting with each other when appropriate damping is introduced, resulting in an exceptional point (EP) degeneracy.

The most common source of damping in structures is Coulomb friction, which opposes the relative motion between surfaces. A coupled oscillators system with frictional damping is shown in \Cref{her2}(a). Two oscillators of masses $m_1$ and $m_2$ and Hookean springs of stiffnesses $k_1$ and $k_2$, respectively are mounted on rigid walls and coupled together by another Hookean spring of stiffness $\kappa$. Frictional forces of magnitudes $\gamma_1 N_1$ and $\gamma_2 N_2$ are resisting the motion of $m_1$ and $m_2$ respectively. Where $\gamma_1, \gamma_2$ are coefficients of friction and $N_1, N_2$ are the contact forces. The governing equations of motion for the oscillators can be written as

\begin{equation}\label{eq1}
	m_1\ddot{x_1}+k_1x_1+(\gamma_1 N_1)\times sgn(\dot{x_1})+\kappa (x_1-x_2)=0
\end{equation}
\begin{equation}\label{eq2}
	m_2\ddot{x_2}+k_2x_2+(\gamma_2 N_2)\times sgn(\dot{x_2})+\kappa (x_2-x_1)=0
\end{equation}

where $x_1(t)$ and $x_2(t)$ are the instantaneous positions of $m_1$ and $m_2$ oscillators respectively, and $sgn$ is the signum function, e.g., $ sgn\left(\dot{x}_{1}\right)=\frac{\dot{x}_{1}}{\left|\dot{x}_{1}\right|}$. Considering steady state solutions for $x_1$ and $x_2$ i.e., $x_1(t)=X_1e^{i\omega t}$ and $x_2(t)=X_2e^{i\omega t}$ which gives $ sgn\left(\dot{x}_{1}\right)=\frac{i\omega X_1}{\left|i\omega X_1\right|}=i=\sqrt{-1}$, where, $X_1$ and $X_2$ are the complex amplitudes of vibration of the two individual oscillators. Also considering $N_1=k_1x_1$ and $N_2=k_2x_2$, which means that the friction force is due to the structural (or hysteretic) damping property of the spring's constituent material. Unlike viscous damping, structural damping is rate-independent. It is usually found in metals, where it arises due to crystal defects and internal friction from grain boundary sliding.

Assuming $m_1=m_2=m$ and $k_1=k_2=k$ for symmetry, and $\omega_0=\sqrt{k\over m}$, $\Omega = {\omega\over {\omega_0}}$, and $\bar{\kappa}={\kappa\over k}$ for making the parameters dimensionless, final equations of motion in matrix form can be written as

\begin{equation}\label{eq3}
    \left[\begin{array}{cc}
-\Omega^{2}+1+i \gamma_{1}+{\bar{\kappa}} & -\bar{\kappa} \\
-\bar{\kappa} & -\Omega^{2}+1+i \gamma_{2}+\bar{\kappa}
\end{array}\right]\left\{\begin{array}{l}
X_{1} \\
X_{2}
\end{array}\right\}=0
\end{equation}

\begin{equation}\label{eq4}
\left[\mathcal{H}\right] \left\{\psi \right\}=0
\end{equation}

For non-trivial solutions of eigenvalues $(\Omega)$ and eigenvectors $\{\psi\}$, substituting the determinant of $[\mathcal{H}]$ equal to 0 yields,
\begin{equation}\label{eq5}
\Lambda(\Omega)=\Omega^{4}-\left(2+i(\gamma_1+\gamma_2)+2 \bar{\kappa}\right) \Omega^{2}+1+i\left(\gamma_{2}+\gamma_{1}\right)+i\left(\gamma_{1}+\gamma_{2}\right) \bar{\kappa}+2 \bar{\kappa}-\gamma_{1} \gamma_{2}=0
\end{equation}

\begin{figure}[t]
	\centering
	\includegraphics[width=0.5\textwidth]{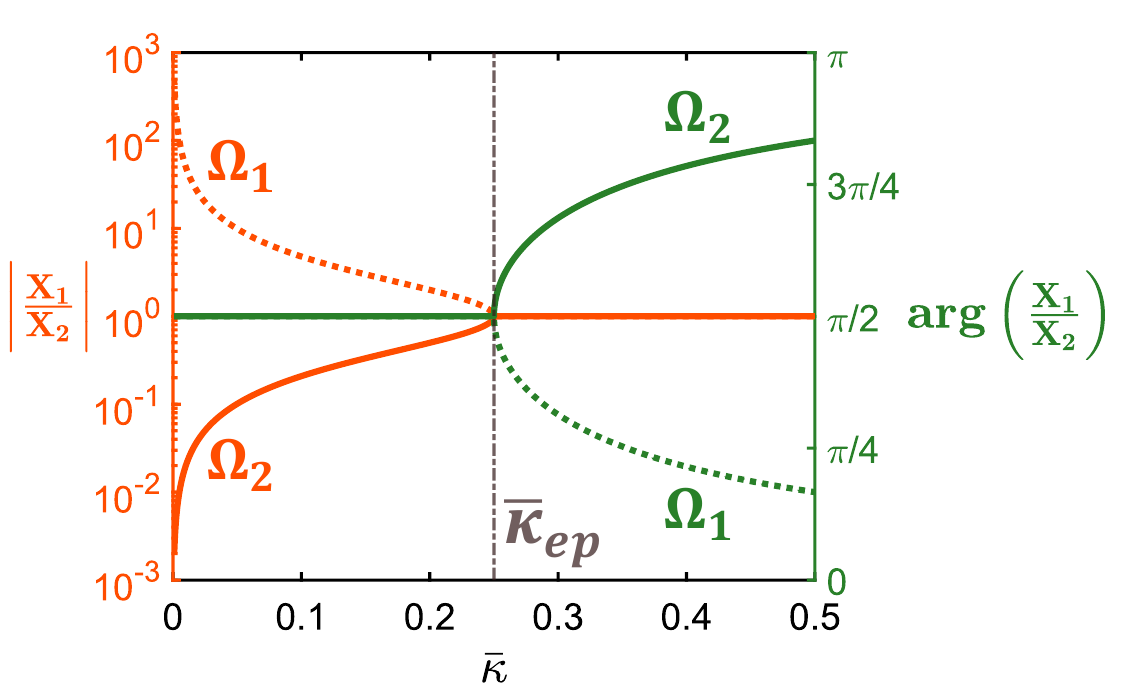}
	\caption{(a) Magnitudes and phases of eigenvectors corresponding to both natural frequencies of coupled oscillators dimer with differential structural damping $(\gamma_1=0,\;\gamma_2=0.5)$.}
	\label{her5}
\end{figure}

Solving the above equation yields the two complex natural frequencies of the system, $\Omega_1$ and $\Omega_2$. In \Cref{her2}, we plot the real and the imaginary parts of $\Omega_1$ and $\Omega_2$ as functions of normalized coupling $(\bar{\kappa})$ for both PT-symmetric (\Cref{her2}(b)) and passive non-Hermitian (differential loss only) cases (\Cref{her2}(c)). To balance the gain and the loss in the PT-symmetric case, we consider $\gamma_1=-\gamma$ and $\gamma_2=\gamma=0.25$. As the coupling increases, an EP emerges at $\bar{\kappa}_{ep}=\gamma$ (refer to \hyperref[section:sd]{SI} for derivation). In the passive non-Hermitian case with no gain, we consider the simplest differential-loss condition (zero loss-loss) by assuming $\gamma_1=0$ and $\gamma_2=\gamma=0.5$. In such a system, the EP appears at $\bar{\kappa}_{ep}={\gamma \over2}$ (refer to \hyperref[section:sd]{SI} for derivation). Noteworthy is the sharp and orthogonal bifurcation at the EP. Here the sharp bifurcation characterizes the zero gap between the complex natural frequencies at the EP i.e., $\left|\Omega_1-\Omega_2\right|=0$. The orthogonal bifurcation is given by the $\tan^{-1}\left(\partial \Omega \over \partial \bar{\kappa}\right)\approx \pi/2$ at the EP---which implies that an infinitesimal variation in $\bar{\kappa}$ will result in a large splitting of frequencies (i.e. $\Omega_2$-$\Omega_1$), suitable for highly sensitive EP-based sensors. We also observe a similar bifurcation in corresponding eigenvectors $\psi=\left[X_1/X_2,1\right]$ (\Cref{her5}), a hallmark feature of EPs. At the EP, the magnitude of eigenvectors $\left|X_1 \over X_2\right|\approx 1$ and phase difference $arg\left(X_1 \over X_2\right) \approx \pi / 2$ for both the modes, which suggest that the modes are degenerate. Conclusively, we have shown a viable pathway to realizing an EP degeneracy in a passive (with no gain) physical system with differential loss. Despite the simplicity, the above system with frictional differential damping exhibits a sharp EP degeneracy in a $2\times 2$ Hamiltonian system similar to coupled optical waveguides \citep{ruter2010observation}, LCR circuits \citep{schindler2011experimental}, acoustic cavities \citep{ding2016emergence}, microwave cavities \citep{peng2014parity}, and quantum harmonic oscillators \citep{bender2002complex}.  

\section{Viscoelastic solids as dissipative elements}

While structural damping is the simplest model for material damping, it is not the most common type of damping present in engineering materials that are used in applications requiring damping. Usually, engineering materials exhibit viscous damping, which is intrinsically connected to materials' elastic properties, making the materials viscoelastic \citep{lakes2017viscoelastic,findley2013creep}. In contrast to structural damping, where the stiffness and damping parameters $(k_1,k_2,\gamma_1,\gamma_2)$ are constants, viscoelastic materials are rate dependent. Viscoelastic materials distinctively exhibit a time decaying relaxation modulus $(E(t)=E_{\infty}+E_t(t))$ whose Fourier transform gives a complex dynamic modulus with frequency dependent real and imaginary parts $(E_d(\omega)=E'(\omega)+iE''(\omega))$. The real part $E'(\omega)$ is called the storage modulus, which is associated with the amount of elastic energy that the material stores under deformation. The imaginary part $E''(\omega)$ is called the loss modulus, which is associated with the energy that the material dissipates under cyclic loading. The ratio $\tan (\delta) = {E''(\omega) \over E'(\omega)}$, known as the loss-tangent, is a quantitative measure of the damping capacity of a viscoelastic material.

To realize the non-Hermitian metamaterial, we replace the structurally damped spring of the coupled oscillators model with a viscoelastic material of dynamic modulus $E_d(\omega)$, cross-sectional area $A$, and length $L$ (\Cref{her3}(a)). The resultant dynamic stiffness of the material is given by $k_d(\omega)=E_d(\omega){A \over L}$. The modified governing equations for the non-Hermitian metamaterial with viscoelastic damping elements in frequency domain can be written as,

\begin{equation}\label{eq6}
	-\Omega^2{X_1}+X_1+\bar{\kappa} (X_1-X_2)=0
\end{equation}
\begin{equation}\label{eq7}
	-\Omega^2{X_2}+{k_d(\Omega) \over k} X_2+\bar{\kappa} (X_2-X_1)=0
\end{equation}

where $X_1$ and $X_2$ are complex amplitudes, $\bar{\kappa}={\kappa \over k}$, $\Omega={\omega \over \omega_0}$, $\omega_0=\sqrt {k\over m}$, and $k$ is the stiffness of Hookean spring (non-lossy) as shown in \Cref{her3}(a).

We investigate three different models of viscoelastic materials in comparison: the Kelvin-Voigt (KV) model, the Standard Linear Solid (SLS), and the Kelvin-Voigt fractional derivative (KVFD) model. The relaxation moduli and the corresponding frequency-dependent dynamic moduli of the three viscoelastic materials are listed in \Cref{tab}.

\begin{table}[]
\centering
\caption{Relaxation and dynamic moduli of KV solid, SLS and KVFD solid.}
\resizebox{\textwidth}{!}{
\begin{tabular}{cccc}
\hline
                                             & \textbf{Kelvin-Voigt (KV)} & \textbf{Standard Linear Solid (SLS)} & \textbf{Kelvin-Voigt fractional derivative (KVFD} \\ \hline
\multirow{2}{*}{\textbf{Relaxation Modulus}} & \multirow{2}{*}{\Large{$E_{\infty}+\eta \hat{\delta}(t)$}}         & \multirow{2}{*}{\Large{$E_{\infty}+E_{t} e^{-\frac{t}{\tau}}$}}                   & \multirow{2}{*}{\Large{$E(t)=E_{\infty}+\eta \frac{1}{\Gamma\left(1-\alpha\right)} t^{-\alpha}$}}                                \\
                                             &                            &                                      &                                                   \\ \hline
\multirow{2}{*}{\textbf{Dynamic Modulus}}    & \multirow{2}{*}{\Large{$E_{\infty}+i\omega \eta$}}         & \multirow{2}{*}{\Large{$E_{\infty}+E_t{\omega^2 \tau^2 \over 1+\omega^2\tau^2}+iE_t{\omega \tau \over 1+\omega^2\tau^2}$}}                   & \multirow{2}{*}{\Large{$E_{\infty}+\eta \omega^{\alpha}\left(\cos \left(\frac{\pi}{2} \alpha\right)+i \sin \left(\frac{\pi}{2} \alpha\right)\right)$} }                                \\
                                             &                            &                                      &                                                   \\ \hline
\end{tabular}}\label{tab}
\end{table}

\subsection{Kelvin-Voigt (KV) materials}

\begin{figure}[t]
	\centering
	\includegraphics[width=1\textwidth]{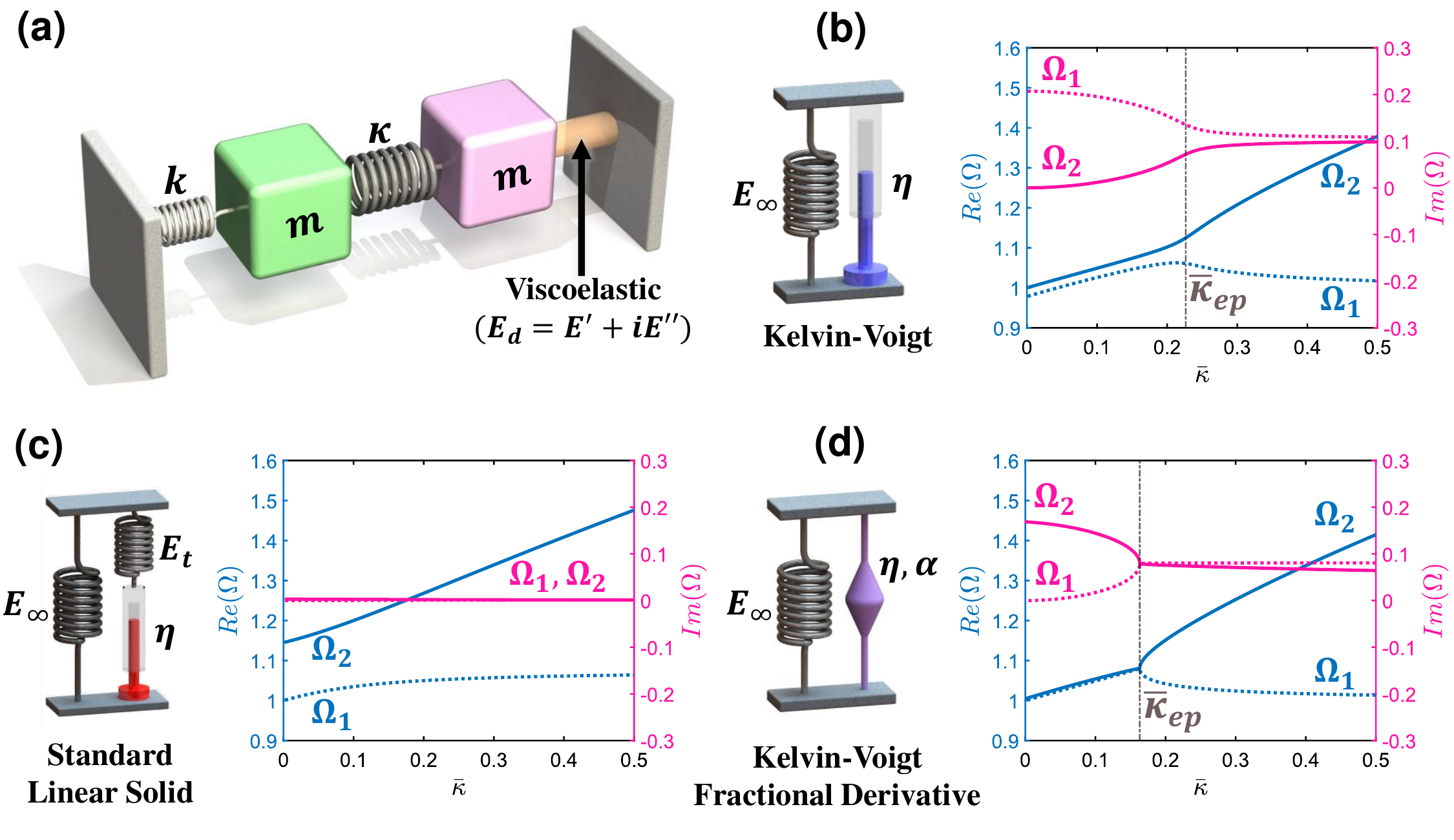}	\caption{(a) Coupled oscillators dimer with differential damping achieved by a Hookean spring of stiffness $k$ and a viscoelastic material of dynamic modulus $E_d$, cross-sectional area $A$ and length $L$. Physical representations and corresponding plots of real and imaginary parts of natural frequencies for (b) KV solid $(\gamma=0.41)$, (c) SLS $\left( {k_\infty \over k }=0.81, {k_t \over k }=0.50, T=20\pi  \right)$, and (d) KVFD solid $\left({k_\infty \over k }=0.25, {\gamma \over k }=0.80, \alpha=0.25, \omega_0=1315\pi  \right)$ as functions of coupling.}
	\label{her3}
\end{figure}

Kelvin-Voigt is the simplest model of a viscoelastic solid. It consist of a spring and a newtonian dashpot in parallel combination (\Cref{her3}(b)). Under sinusoidal loading, the complex dynamic modulus of KV solid is given by 
\begin{equation}\label{eq8}
    E_d=E_{\infty}+i\omega\eta 
\end{equation}
where, $E_{\infty}$ is the modulus of the spring, $\eta$ is the viscous damping coefficient and $\omega$ is the frequency of the dynamic load. Notably, KV materials exhibit a frequency-independent storage modulus $(E_{\infty})$ and a loss modulus $(\omega\eta)$ that is linearly dependent on frequency. Assuming $k=E_{\infty}{A\over L}$ for symmetry in real part of stiffness and ${A\over L}{\eta\over m \omega_0}=\gamma$ as the dimensionless damping coefficient, the governing equations can be written in matrix form as follows

\begin{equation}\label{eq9}
    \left[\begin{array}{cc}
-\Omega^{2}+1+i\Omega \gamma_1+{\bar{\kappa}} & -\bar{\kappa} \\
-\bar{\kappa} & -\Omega^{2}+1+i \Omega \gamma_2+\bar{\kappa}
\end{array}\right]\left\{\begin{array}{l}
X_{1} \\
X_{2}
\end{array}\right\}=0
\end{equation}

\begin{figure}[t]
	\centering
	\includegraphics[width=0.5\textwidth]{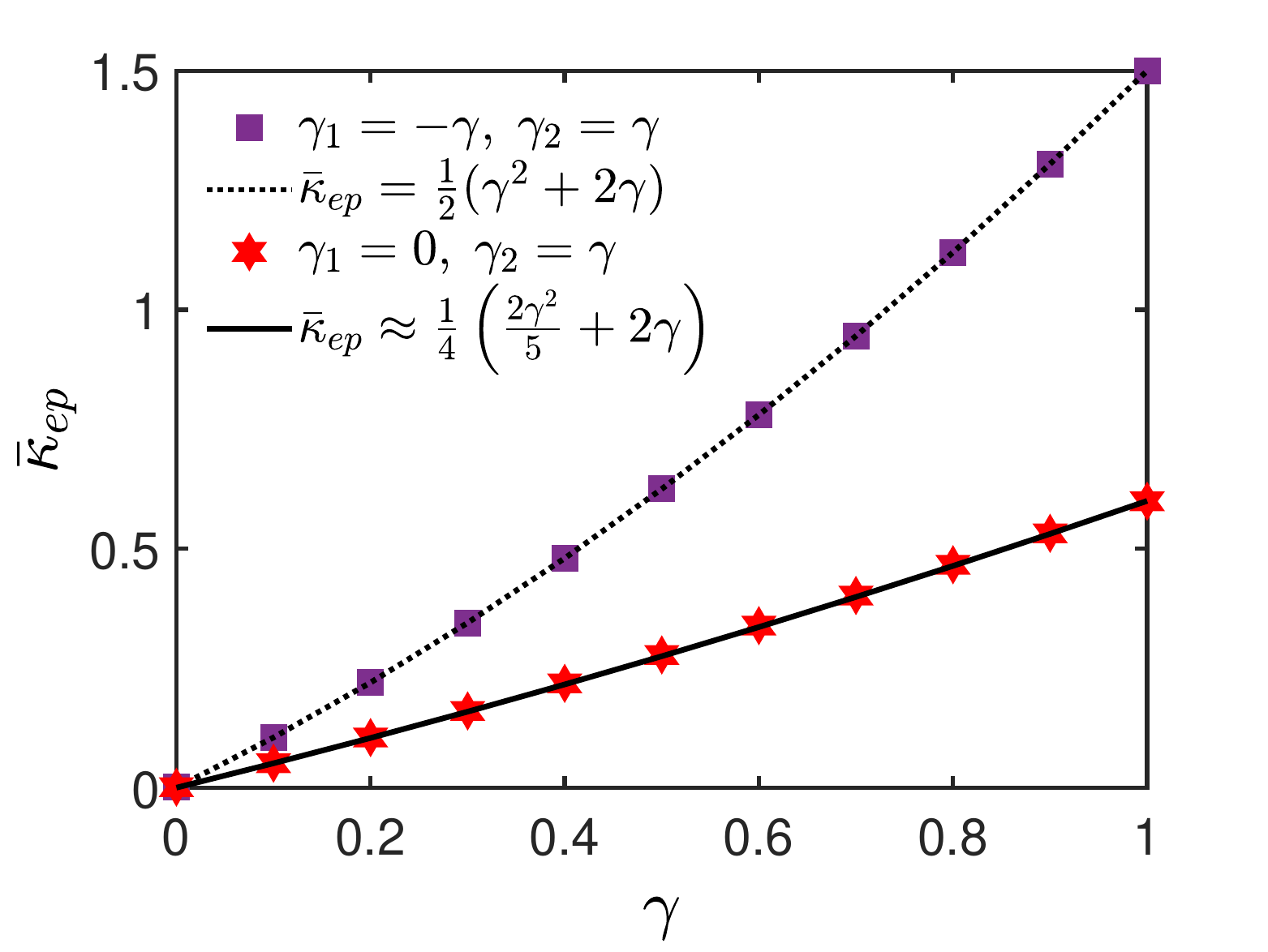}	\caption{(a) Comparison of $\bar{\kappa}_{ep}$ as a function of $\gamma$ between PT-symmetric and the passive non-hermitian case for coupled oscillators with KV material as the loss and gain element.}
	\label{her9}
\end{figure}

\begin{figure}[t]
	\centering
	\includegraphics[width=0.5\textwidth]{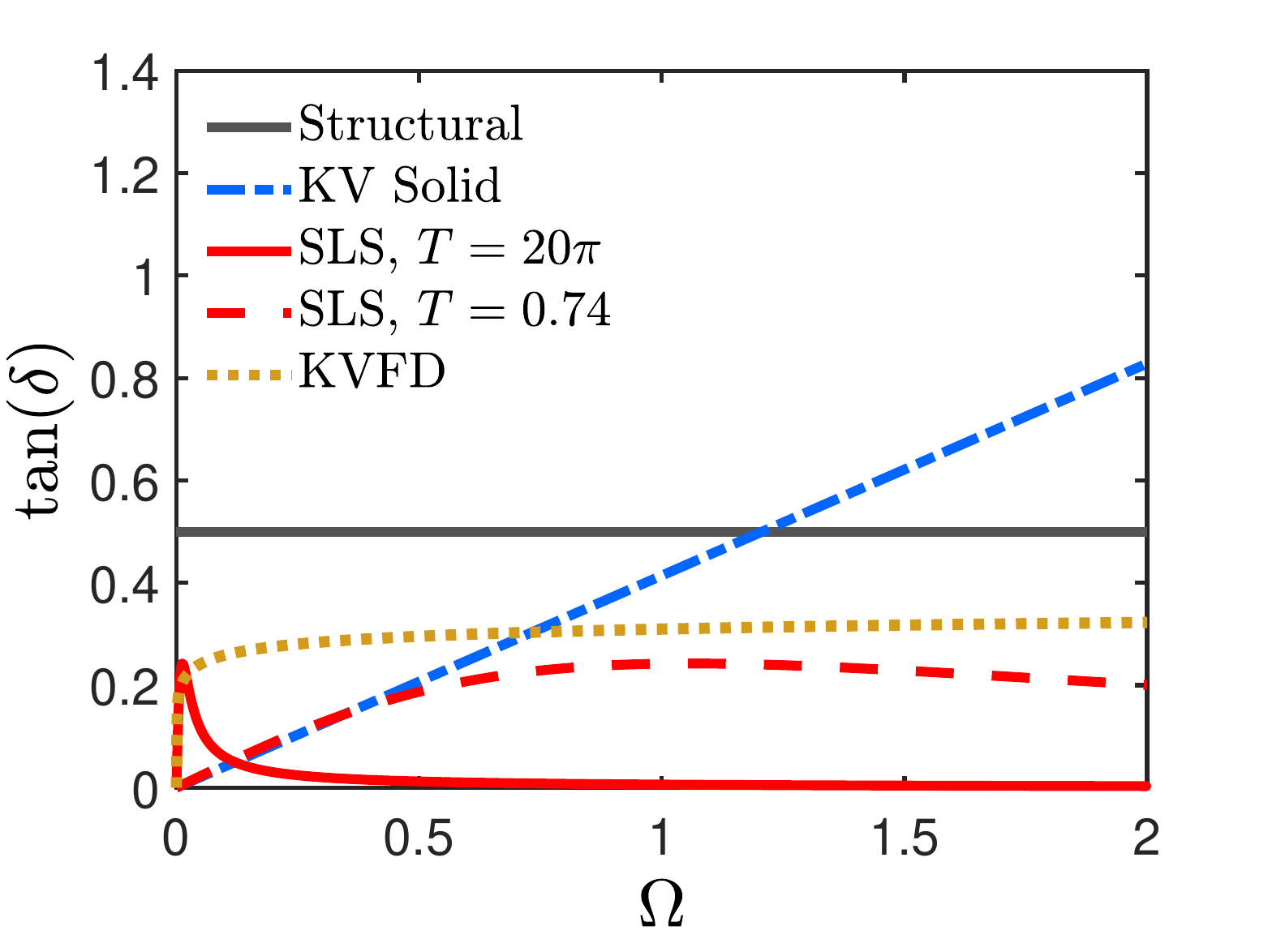}	\caption{(a) Loss tangent $(\tan (\delta))$ as a function of frequency for structural damping $(\gamma=0.5)$, KV solid $(\gamma=0.41)$, SLS $\left( {k_\infty \over k }=0.81, {k_t \over k }=0.50, T=20\pi  \right)$ , SLS $\left({k_\infty \over k }=0.81, {k_t \over k }=0.50, T=0.74  \right)$ and KVFD solid $\left({k_\infty \over k }=0.25, {\gamma \over k }=0.80, \alpha=0.25, \omega_0=1315\pi  \right)$}
	\label{her6}
\end{figure}

Here, we first consider the PT-symmetric case $(\gamma_1=-\gamma,\gamma_2=\gamma)$. To balance with loss, the gain must also be linearly dependent on frequency. Physical example of such gain is self-excitation of spring-block on a moving conveyor due to velocity-dependent friction coefficient \citep{rao1995mechanical}. On putting the determinant of the Hamiltonian matrix in Eq. \ref{eq9} equal to zero and solving for $\Omega$, we observed a sharp EP at $\bar{\kappa}_{ep}=\frac{1}{2}(\gamma^2+2\gamma)$ (\Cref{her9}) \citep{bender2013twofold}, with eigen frequencies very similar to the PT-symmetric case in structural damping (\Cref{her2}(b)). For the passive non-Hermitian case $(\gamma_1=0,\gamma_2=\gamma)$, the real and imaginary parts of the complex natural frequencies $\Omega_1$ and $\Omega_2$ are plotted as functions of $\bar{\kappa}$ in \Cref{her3}(b). Contrary to the structural damping case, the passive non-Hermitian metamaterial made of a KV material does not exhibit a sharp EP bifurcation. Instead, we observe an exceptional region \citep{bender2013twofold,gu2021controlling} where the modes gradually bifurcate. In the exceptional region, we identify that the gap between $\Omega_1$ and $\Omega_2$, i.e., $\left|\Omega_1-\Omega_2\right|$ is minimum at $\bar{\kappa}_{ep}\approx {1\over 4}\left(\frac{2}{5}\gamma^2+2\gamma\right)$ (\Cref{her9}), which we define as the location of a fictitious EP $(\bar{\kappa}_{ep})$ as shown in \Cref{her3}(b). We hypothesize that the indistinct EP transition observed here is due to a strong linear dependence of $\tan (\delta)$ on the frequency, i.e., to achieve a sharp EP bifurcation, $\tan (\delta)$ should converge to a constant value as the frequency is increased (\Cref{her6}). Strong frequency dependence of $\tan (\delta)$ in KV solid is associated with dirac delta function $(\hat{\delta}(t))$ in transient part of relaxation modulus (\Cref{tab}), which indicates that material relaxes abruptly at $t=0$ and does not relax further afterwards $(\hat{\delta}(t>0)=0)$. To test our hypothesis of the effects of linearly dependent $\tan (\delta)$ on the frequency, we next investigate SLS and KVFD viscoelastic solids, which undergo stress relaxation for longer duration.

\subsection{Standard Linear Solid (SLS) materials}

The SLS is a viscoelastic material model consisting of a spring connected in parallel to a Maxwell element (a spring and a dashpot in series) as shown in \Cref{her3}(c). The relaxation modulus of SLS decays exponentially with time, and corresponding storage and loss moduli are both frequency-dependent (\Cref{tab}). Assuming $E_{\infty}$ and $E_t$ as the moduli of springs, $\eta$ as the damping coefficient of the dashpot and $\tau={\eta\over E_t}$ as the relaxation time, the dynamic modulus can be written as,

\begin{equation}\label{eq10}
{E_d(\omega)} =  E_{\infty}+E_t {\omega^2\tau^2 \over {1+\omega^2\tau^2}}+iE_t {\omega \tau \over {1+\omega^2\tau^2}}
\end{equation}

Assuming $k_{\infty}=E_{\infty}{A\over L}$, $k_{t}=E_{t}{A\over L}$, $\Omega={\omega \over \omega_0}$, $\omega_0=\sqrt{k \over m}$, and $T=\omega_0 \tau$ (dimensionless relaxation time), the normalized dynamic stiffness can be written as,

\begin{equation}\label{eq11}
{k_d(\Omega)\over k} = {1\over k} \left( k_{\infty}+k_t {\Omega^2T^2 \over {1+\Omega^2T^2}}+ik_t {\Omega T \over {1+\Omega^2T^2}}    \right)
\end{equation}

\begin{figure}[!htb]
	\centering
	\includegraphics[width=0.5\textwidth]{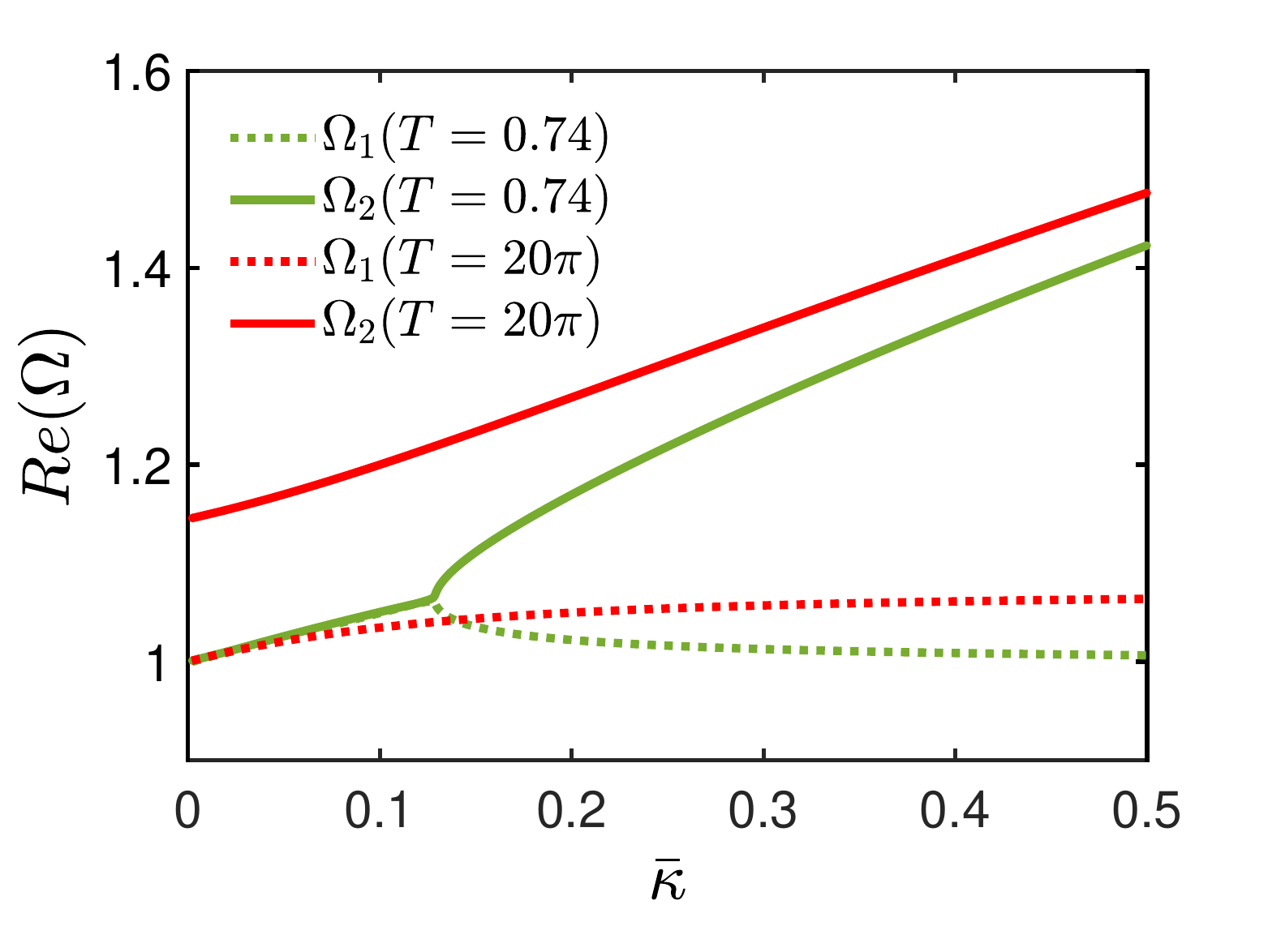}
	\caption{Real part of natural frequencies for $T=0.74$ and $T=20\pi$}
	\label{her9A}
\end{figure}

Since it is physically not possible to create gain and maintain symmetry in materials systems with the real part of modulus also frequency dependent, from here onwards, we only investigate passive non-hermitian systems with differential damping (noLoss-Loss). On substituting Eq. (\ref{eq11}) in Eqs. (\ref{eq6},\ref{eq7}) and solving for $\Omega$, we obtain two resonant frequencies $\Omega_1$, $\Omega_2$. The real and imaginary parts of $\Omega_1$ and $\Omega_2$ are plotted in \Cref{her3}(c). In this case, we didn't observe any EP formation for all reasonable values of material parameters. An EP only starts to appear when $T$ is very small $(T < 1)$, which is physically not possible. In \Cref{her6}, comparison of $\tan (\delta)$ for $T=20\pi$ and $T=0.74$ indicates that, for shorter relaxation times, the $\tan (\delta)$ has a plateau like regime and is almost constant for $\Omega\approx1$ resulting in the formation of an exceptional point (\Cref{her9A}). However, for longer relaxation times, the $\tan (\delta)$ quickly decays to almost zero, making the imaginary part of the dynamic modulus so small to show any effect. Supposedly, a material model representing a weighted sum of multiple relaxation times---both short and long---will resolve this issue. Usually, experimental measurements of relaxation modulus of viscoelastic materials are best described by a prony series, a sum of exponentials physically represented by the generalized Maxwell (GM) model \citep{babaei2016efficient,jalocha2015revisiting,diani2012predicting}. The GM model is similar to SLS model with multiple Maxwell elements connected in parallel, resulting in a discrete spectrum of relaxation times \citep{lakes2017viscoelastic,findley2013creep}. Relaxation modulus $E(t)$ and dynamic modulus $E_d(\omega)$ for the GM model can be written as,

\begin{equation}\label{eq12}
    E(t)=E_{\infty}+\sum_{m=1}^{M} E_{m} e^{-\frac{t}{\tau_{m}}}
\end{equation}

\begin{equation}\label{eq13}
    E_d(\omega)=E_{\infty}+\sum_{m=1}^{M} \left( E_{m} {\omega^2\tau_m^2 \over {1+\omega^2\tau_m^2}} +i E_m {\omega \tau_m \over {1+\omega^2\tau_m^2}} \right)
\end{equation}

However, fitting these many terms to experimental data is neither practical nor desirable. So, a power law relaxation function $(E(t)=E_{\infty}+E_{\alpha} t^{-\alpha})$ is generally used for materials which relaxes over a large spectrum of relaxation times \citep{schiessel1993hierarchical,kelly2009fractal,mainardi2010fractional,xiao2016equivalence}. For sufficient number of terms, the prony series (\ref{eq12}) is approximately equivalent to a shifted power law relaxation function $E(t)=E_{\infty}+E_{\alpha}(t+t_l)^{-\alpha} \; (t_l \geq 0,\;1>\alpha>0)$ as follows (more details in \hyperref[section:sd3]{Appendix B}),

\begin{equation}\label{eq14}
E_{\infty}+E_{\alpha} (t+t_l)^{-\alpha} = E_{\infty}+\frac{E_{\alpha}}{\Gamma(\mathrm{\alpha})} \int_{0}^{\infty}\left(e^{-\frac{t_{l}}{\tau}} \frac{1}{\tau^{\alpha+1}}\right) e^{-\frac{t}{\tau}}d \tau
\end{equation}

where $\Gamma$ is the gamma function. Eq. (\ref{eq14}) can be converted into an approximate discrete sum using trapezoidal rule as follows

\begin{equation}\label{eq15}
E_{\infty}+E_{\alpha} (t+t_l)^{-\alpha} \approx E_{\infty}+\sum_{m=1}^{M\to \infty} E_m e^{-\frac{t}{\tau_{m}}}
\end{equation}

\begin{equation}\label{eq16}
E_{m}=\frac{E_{\alpha}}{\Gamma(\alpha)}\left\{\begin{array}{lr}
0.5 \times e^{-\frac{t_{l}}{\tau_{m}}}\left(\tau_{m}\right)^{-(\alpha+1)} \times\left(\tau_{m+1}-\tau_{m}\right) & m=1 \\
0.5 \times e^{-\frac{t_{l}}{\tau_{m}}}\left(\tau_{m}\right)^{-(\alpha+1)} \times\left(\tau_{m+1}-\tau_{m-1}\right) & 1<m<M \\
0.5 \times e^{-\frac{t_{l}}{\tau_{m}}}\left(\tau_{m}\right)^{-(\alpha+1)} \times\left(\tau_{m}-\tau_{m-1}\right) & m=M
\end{array}\right.
\end{equation}

where, $\tau_m=\tau_1,\tau_2,\tau_3.......\tau_M$ are relaxation times and $E_m$ are corresponding coefficients of a discrete relaxation spectrum $\hat{H}(\tau)$ given by \citep{lakes2017viscoelastic}
\begin{equation}\label{eq17}
\hat{H}(\tau)=\sum_{m=1}^{M} \tau E_{m} \hat{\delta}\left(\tau-{\tau_{m}}\right)
\end{equation}

Eqs. (\ref{eq15}, \ref{eq16}) suggests that instead of using prony series, the simpler power-law relaxation modulus can be used to model realistic viscoelastic materials \citep{craiem2007fractional,gomez2022experimental,bagley1983theoretical,mainardi2010fractional}. Physical representation of power-law relaxation is the Kelvin-Voigt fractional derivative (KVFD) model. Thus, in the next section, we investigate the possibility of achieving an EP using KVFD solid as the non-Hermitian element.

\begin{figure}[t]
	\centering
	\includegraphics[width=\textwidth]{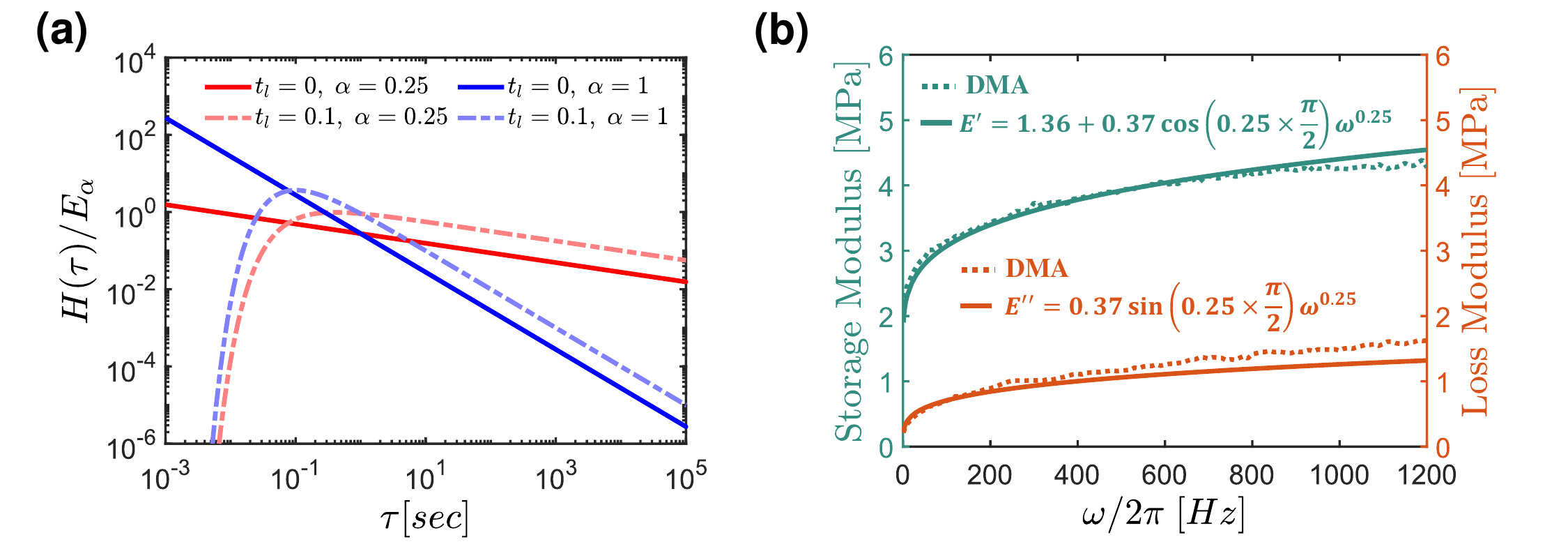}	\caption{(a) Normalized relaxation spectrums for different combinations of $t_l$ and $\alpha$. (b) Experimentally measured storage and Loss modulus of Polydimethylsiloxane (PDMS) as functions of excitation frequency by performing the dynamic mechanical analysis (DMA). Best-fit power-law curves indicate that PDMS follows the KVFD model.}
	\label{her8}
\end{figure}

\subsection{Kelvin-Voigt fractional derivative (KVFD) materials}

The KVFD solid is similar to the KV solid with a springpot replacing the dashpot. Springpot represents an interpolation between a spring and a dashpot. For an applied strain $\epsilon$, the stress response of a springpot will be $\sigma = \eta {d^\alpha\epsilon \over d t^\alpha}$, where, $0\leq\alpha\leq1$, with $\alpha=1$ indicating a dashpot and $\alpha=0$ indicating a spring. The KVFD model we have consider here consists of a sprintpot connected in parallel to a spring (\Cref{her3}(d)). Usually one or two springpots in parallel are sufficient to accurately model both transient and dynamic behavior of a viscoelastic material \citep{craiem2007fractional}.

The relaxation modulus of KVFD solid obeys a power-law decay as follows, 
\begin{equation}
E(t)=E_{\infty}+E_{\alpha}t^{-\alpha}\;,\;\;\;E_{\alpha}=\frac{\eta}{\Gamma(1-\alpha)}.  
\end{equation}

Sometimes, a more general function---shifted power law $(E(t)=E_{\infty}+E_{\alpha}(t+t_l)^{-\alpha})$ is used to avoid the divergence of the transient term at $t=0$. From Eq. (\ref{eq14}), we can extract the relaxation spectrum $H(\tau)$ as follows

\begin{equation}\label{eq19}
E_{\infty}+E_{\alpha} (t+t_l)^{-\alpha} = E_{\infty}+ \int_{0}^{\infty} H(\tau)\; e^{-\frac{t}{\tau}}\;\frac{d\tau}{\tau}\;,\;\;\;H(\tau)= \frac{E_{\alpha}}{\Gamma(\alpha)} e^{-\frac{t_l}{\tau}} \frac{1}{\tau^\alpha}
\end{equation}

\begin{figure}[t]
	\centering
	\includegraphics[width=0.5\textwidth]{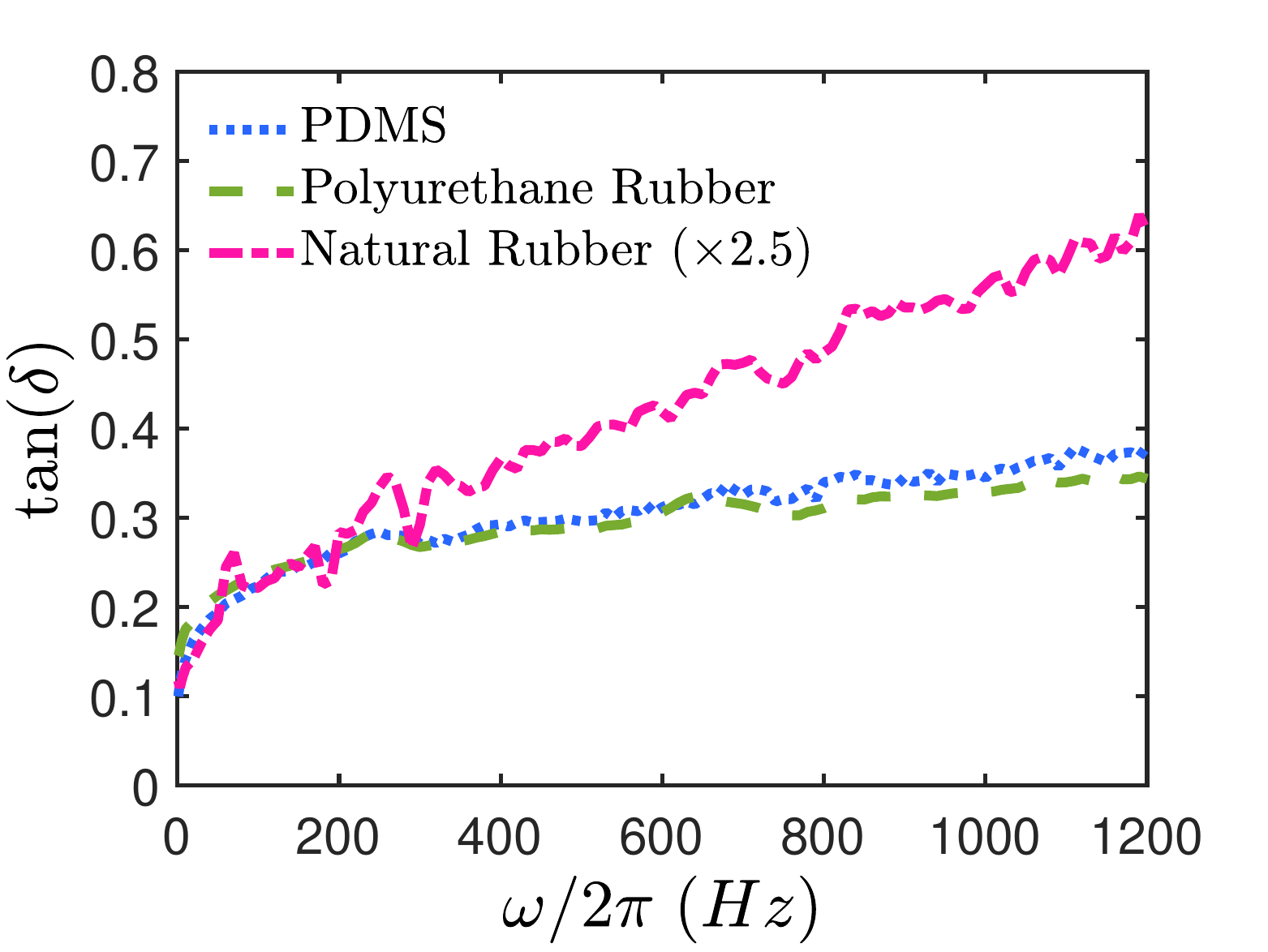}	\caption{Comparison of frequency-dependent loss-tangent between PDMS, Polyurethane rubber, and Natural rubber. To provide a better comparison, we scale the loss tangent of natural rubber by 2.5 times.}
	\label{tand}
\end{figure}

In contrast to GM model (\ref{eq17}), relaxation spectrum for power law model (\ref{eq19}) is continuous. In \Cref{her8}(a), the normalized relaxation spectrum of the power law $(H(\tau)/E_{\alpha})$ is plotted as a function of relaxation time for different combinations of $\alpha$ and $t_l$. For $\alpha=1$ and $t_l=0$ (KV solid), shorter relaxation times are more dominant as compared to longer, which is the reason why KV solid undergoes an abrupt relaxation at $t=0$. For $t_l>0$ (shifted power law), the relaxation spectrum peaks at $\tau=t_l/\alpha$, but decays in both directions from the peak. For $t_l=0$ and $\alpha=0.25$ (KVFD), both shorter and longer relaxation times dominate. The effect of such relaxation behavior can be observed in the loss-tangent (\Cref{her6}), which stays almost constant for higher frequencies, making KVFD materials a potential choice for achieving EPs.

The normalized dynamic stiffness (\Cref{tab}) for the KVFD solid shown in \Cref{her3}(d) can be written as,

\begin{equation}\label{eq18}
{k_d(\Omega)\over k} = {1\over k} \left( k_{\infty}+\gamma \Omega^{\alpha} \cos \left({{\pi \over 2}\alpha} \right)+ i\gamma \Omega^{\alpha} \sin \left({{\pi \over 2}\alpha} \right) \right)
\end{equation}

where $k_{\infty}=E_{\infty}{A\over L}$, $\gamma=\eta {\omega_0}^\alpha {A\over L}$, $\Omega={\omega \over \omega_0}$ and $\omega_0=\sqrt{k \over m}$. Substituting Eq. (\ref{eq18}) in Eqs. (\ref{eq6},\ref{eq7}) and solving for $\Omega$ gives two converging solutions, $\Omega_1$ and $\Omega_2$. \Cref{her3}(d) shows the real and imaginary parts of both natural frequencies undergoing a sharp EP bifurcation. The characteristics of bifurcation near the EP for KVFD material are similar to the structural damping case described earlier. Analogous to strcutural damping, for KVFD, $tan(\delta)$ is almost constant and coverges to $\tan \left({\pi \over 2}\alpha \right)$ as $\Omega \to \infty$. This verifies our hypothesis that for a sharp EP bifurcation to occur, the $\tan (\delta)$ should not diverge with frequency and should stay almost constant in the frequency range of interest. Interestingly, compared to structural damping, which is rarely observed in engineering materials, many soft polymers and biological materials exhibit viscoelastic behavior that is best described by fractional derivative models. \Cref{her8}(b) shows the storage and loss modulus of PDMS (Sylgard 184, 1:20 ratio of curing agent to liquid elastomer) that we measured experimentally by performing the dynamic mechanical analysis (DMA). It strongly follows the KVFD model with parameters $\alpha=0.25$, $E_\infty=1.36$, $\eta=0.37$. 

In \Cref{tand}, we compare the frequency-dependent loss tangent of PDMS, polyurethane rubber, and natural rubber that we measured using DMA. PDMS and polyurethane exhibit a slow dependence of $\tan(\delta)$ on the frequency, which is appropriate for forming a sharp exceptional point, whereas natural rubber has a strong dependence on the frequency, which makes it unsuitable. Thus, using elastomers like PDMS and polyurethane rubber in passive non-Hermitian systems makes the realization of sharp EPs experimentally viable \citep{gupta2022reconfigurable}.

 \begin{figure}[t]
	\centering
	\includegraphics[width=\textwidth]{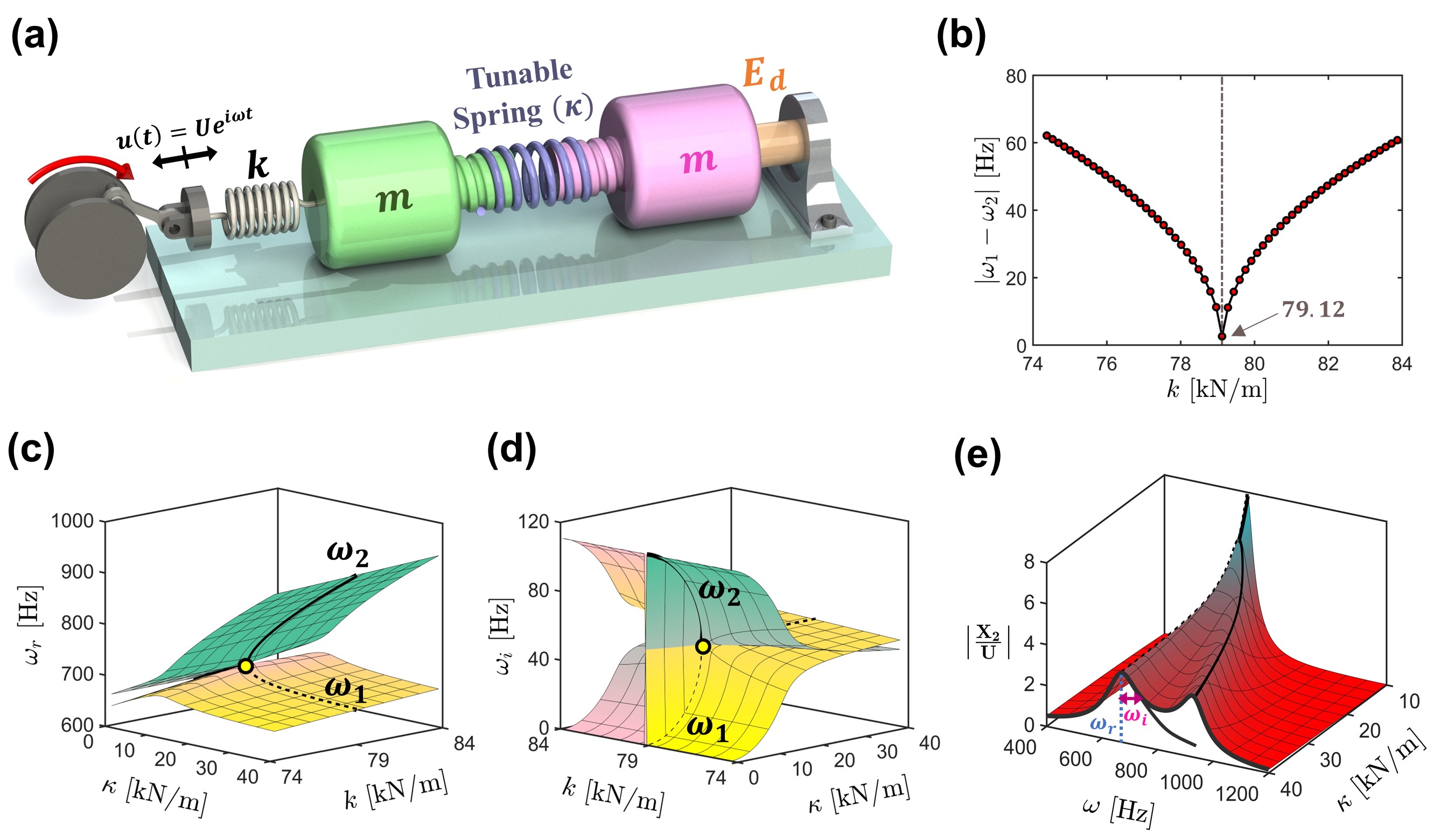}
	\caption{(a) Coupled oscillators with driven left boundary by a sinusoidal actuation. The oscillators are coupled together by a spring of tunable stiffness. (b) Absolute gap between eigen frequencies at the EP is minimized by selecting a Hookean spring of optimal stiffness. (c, d)  Real $(\omega_r)$ and imaginary parts $(\omega_i)$ of natural frequencies $(\omega_1,\omega_2)$ are plotted against coupling for various values of $k$. A sharp exceptional point appears for optimized value of $k$. (e) For the optimized system, normalized forced vibration amplitude of second oscillator is plotted as a function of frequency for a range of values of coupling. At each coupling, real and imaginary parts of eigen frequencies can be estimated by fitting the resonance pulse and identifying its peak position and pulse width at $1 /\sqrt{2}$ of pulse height respectively.}
	\label{her7}
\end{figure}


In experiments, the real and imaginary parts of eigenfrequencies can be characterized by measuring the frequency response of the system under sinusoidal excitation. \Cref{her7}(a) shows our model system under forced vibration with the left boundary being driven by applying a sinusoidal displacement $u(t)=Ue^{i\omega t}$. The two oscillators are coupled together by a tunable spring whose stiffness can be adjusted to change the coupling. The required differential-damping is achieved by a Hookean spring of stiffness $k$ (no-Loss) and a PDMS cylinder of area $A$, length $L$, and dynamic modulus $E_d$ (Loss) (\Cref{her8}(b)). 

We assume, $m=4.635\;g$, $A=28.27\; mm^2$ ($6\;mm$ diameter) and $L=1.95 \;mm$. For the known dynamic stiffness of PDMS $(k_d=\frac{E_dA}{L})$, we estimated the corresponding required optimal stiffness of Hookean spring $(k)$ by minimizing the gap between the resonant frequencies $(\left|\omega_1-\omega_2\right|)$ such that the bifurcation at EP is sharpest (\Cref{her7}(b)). Theoretically calculated, real $(\omega_r)$ and imaginary parts $(\omega_i)$ of resonant frequencies are plotted in \Cref{her7}(c, d) in the parameter space of $\kappa$ and $k$. For $k=79.12\; kN/m$, the gap between the two natural frequencies approaches almost zero at the EP, which is an indication of a sharp bifurcation. \Cref{her7}(e) shows the theoretical frequency response of the second oscillator, a sharp bifurcation of a single narrow resonant peak at low couplings $(\kappa<\kappa_{ep})$ into two broad peaks of same linewidth at higher couplings $(\kappa>\kappa_{ep})$. From the frequency response, the real part of resonant frequencies can be estimated from the position of resonant peaks and the imaginary part by measuring the half-width of the peak at $1/\sqrt{2}$ of peak height (\Cref{her7}(e)).

\section{Conclusion}

In this work, we investigated the viability of achieving sharp orthogonal EP bifurcations in passive non-Hermitian elastodynamic systems with viscoelastic materials as the dissipative component. Using a coupled oscillators as the model system, we compared different viscoelastic solids with the general rate-independent hysteretic damping to identify the requisites on frequency-dependent dynamic properties for the creation of an elastodynamic metamaterial that support the formation of an EP. We show that for an EP to form, the loss-tangent of viscoelastic material should stay almost constant in the frequency range of operation, a characteristic encoded in the relaxation spectrum of the material. Our work provides a critical outlook to the emerging field of non-Hermitian elastodynamics by using widely available viscoelastic materials for realizing non-Hermiticity. It paves the way for the development of passive non-Hermitian devices for sensing, actuation, and energy-harvesting.

\section*{Acknowledgements}
We acknowledge the financial support from the Solid Mechanics program of the U.S. Army Research Office (award number: W911NF2010160) and the Dynamics, Controls, and System Diagnostics program of the National Science Foundation (NSF-CMMI-1925530). We also acknowledge the support of Dr. Jizhe Cai in fabricating PDMS samples.
\newpage
\section*{Appendix A. Effect of relaxation time in SLS model on eigen frequencies}
\label{section:sd2}

\section*{Appendix B. Equivalence of generalized Maxwell model and fractional derivative model}
Kelvin-Voigt fractional derivative (KVFD) model undergoes power-law stress relaxation in the time-domain, whereas the generalized Maxwell model is a sum of decaying exponentials. A power law decay can be converted to an approx sum of weighted exponentials as follows

Consider the following Laplace transform $\left(\mathcal{L}\left(f(x)\right)\to F(s)\right)$

\begin{equation}
\mathcal{L} \left(\frac{e^{c x} x^{\alpha-1}}{\Gamma(\alpha)}\right)=\frac{1}{(s-c)^{\alpha}}
\end{equation}

\begin{equation}
\frac{1}{(s-c)^{\alpha}}=\frac{1}{\Gamma(\alpha)} \int_{0}^{\infty} e^{c x} x^{\alpha-1} e^{-s x} d x
\end{equation}

Changing variables $s\to t$ and $c\to -t_l$ $(t_l>0)$

\begin{equation}
\left(t+t_{l}\right)^{-\alpha}=\frac{1}{\Gamma(\alpha)} \int_{0}^{\infty} e^{-t_{l} x} x^{\alpha-1} e^{-t x} d x
\end{equation}

Multiplying $E_{\alpha}$ and adding $E_{\infty}$ on both sides

\begin{equation}
E_{\infty}+E_{\alpha}\left(t+t_{l}\right)^{-\alpha}=E_{\infty}+\frac{E_{\alpha}}{\Gamma(\alpha)} \int_{0}^{\infty} e^{-t_{l} x} x^{\alpha-1} e^{-t x} d x
\end{equation}

Changing variable $x\to \frac{1}{\tau}$

\begin{equation}
E_{\infty}+E_{\alpha}\left(t+t_{l}\right)^{-\alpha}=E_{\infty}+\frac{E_{\alpha}}{\Gamma(\alpha)} \int_{0}^{\infty}\left(e^{-\frac{t_{l}}{\tau}} \frac{1}{\tau^{\alpha}}\right) e^{-\frac{t}{\tau}}\left(\frac{1}{\tau} d \tau\right)
\end{equation}

\label{section:sd3}

\section*{Appendix C. Supplementary data}

Supplementary material related to this article can be found
online

\label{section:sd}

\bibliography{mybibfile}

\end{document}